# Structural phase transitions in perovskite BaCeO$_3$ with data mining and first-principles theoretical calculations

Farha Naaz, Manendra S. Chauhan, Kedar Yadav, Surender Singh, Ashok Kumar, and Dasari L. V. K. Prasad[*]

*Department of Chemistry, Indian Institute of Technology, Kanpur 208016, India*
(Date: Nov 30, 2023)

Several neutron diffraction, Raman spectroscopy, and thermoanalytical experiments conducted over decades have revealed that the perovskite-structured BaCeO$_3$ goes through a series of temperature-induced structural phase transitions. However, it has been frequently observed that the number of phases and the sequence in which they appear as a function of temperature differ between experiments. Insofar as neutron diffraction experiments are concern, in the temperature range of 4.2 to 1273 K, four structures are crystallographically well characterized with three transitions: orthorhombic *Pnma* → orthorhombic *Imma* [563 K] → rhombohedral *R*-3*c* [673 K] → cubic *Pm*-3*m* [1173 K], which lately have been reciprocally realized in the studies of polarized Raman spectroscopy. In contrast, thermoanalytical methods such as dilatometry showed multiple singularities corresponding to at-least three more structural phase transitions at around 830 K, 900 K, and 1030 K, in addition to those recorded by neutron studies. In account of these conflicting experimental findings, we computed free energy phase diagram for BaCeO$_3$ polymorphs employing crystal structure data mining in conjunction with first principles electronic structure and phonon lattice dynamics. A total of 34 polymorphs have been predicted, the most stable of which follows the Glazer classification of the perovskite tilt system, and it has been found that a number of these polymorphs are thermodynamically competing with *Pnma* as the temperature rises. In particular, it has been predicted that the orthorhombic *Cmcm* and tetragonal *P*4/*mbm* phases surpass *Pnma* at 666 K and 1210 K, respectively. At any temperature, two alternate tetragonal phases (*P*4$_2$/*nmc* and *I*4/*mcm*) are also found to be 20 to 30 meV less favored than the *Pnma*. While the calculated stability order of the predicted polymorphs is in acceptable agreement with the results of neutron diffraction, the transitions observed in thermoanalytical studies could be ascribed to the development of four novel phases (*Cmcm, P*4/*mbm, P*4$_2$/*nmc,* and *I*4/*mcm*) at intermediate temperatures. However, we analyze that the rhombohedral *R*-3*c* phase predominantly stabilized over a broad temperature field, masking all subsequent phases up until the cubic *Pm*-3*m*. Consequently, the novel phases predicted to occur in thermoanalytical studies are only fleetingly metastable. The calculated phonons additionally demonstrate that the high temperature phases are not quenchable down to room temperature. The theoretical results presented reconcile the apparent inconsistencies observed thus far in the experiments.

---

[*] To whom correspondence should be addressed. Email: dprasad@iitk.ac.in

# INTRODUCTION

BaCeO$_3$ is one of the most well-known perovskite ceramic-oxide that has been extensively studied from the perspective of crystal chemistry, structural phase transitions, and solid oxide electrochemical devices and sensors [1]. In fact, it is one of the few oxides that exhibit relatively high proton conduction at elevated temperatures, and several of its doped variants are also shown to have increased ionic conductivities ($10^{-1}$ to $10^{-2}$ S cm$^{-1}$ at 900 K) [2,3]. Owing to its excellent ionic conductivity and versatile perovskite structure-type that is tolerant to many dopants, the BaCeO$_3$ ceramics are proposed as potential electrolytes for the prospective proton conducting solid oxide fuel cells [4,5].

The solid oxide fuel cells are typically operated between temperatures 800 to 1300 K and in these the cells are saturated with the partial pressures of gaseous components. While certain dopants stabilize or destabilize cerates, the undoped BaCeO$_3$ alone is quite reactive under these fuel cell operating conditions. For example, it has been shown that BaCeO$_3$ is thermodynamically unstable, decomposing into hydroxides, carbonates and cerium oxide in the presence of air, water and carbon dioxide fuel cell atmospheres [6-9]. The scant chemical stability of this cerate at high temperature has shown to hamper its applicability as electrolyte and other redox reactions involved in the electrochemical reactions.

If the proton conduction and chemical stability of the electrolyte were to be understood in making stable and sustainable cerates for high temperature electrochemical and other physicochemical processes and properties, the knowledge of the crystal structure of BaCeO$_3$ at elevated temperatures is fundamentally important. The properties of the materials critically rely upon their crystal structures. To this end, several experiments have been carried out to investigate the structure of BaCeO$_3$ at high temperatures. At STP, although the ground state crystal structure of BaCeO$_3$ was unsettled for decades, to which we will return; at high temperatures until reaching maximum 1500 K, it was reported in various experiments that BaCeO$_3$ undergoes cascades of structural phase transformations [10-12]. These structural phase transitions and their characterization that is the crystal structures and nature of the phase transitions as a function of temperature have been quite comprehensively investigated by several independent researchers employing a variety of techniques such as X-ray or Neutron diffraction, Raman spectroscopy, and thermoanalytical experiments [10-25]. However, besides the fact that in all of these experiments at high temperature BaCeO$_3$ acquires different polymorphic structures, the results of the experiments disagree on a certain number of observations: They differed in their structure assignments, thermodynamic order of phase transition, and the number of phases identified is often found to be contentious. These differences are in general attributed to the nature of samples (twin domains/walls, antiphase boundaries etc.), synthesis methods and the experimental techniques used in discerning the subtle tilts of CeO$_6$ octahedral units that are responsible for the high temperature structural phase transitions in BaCeO$_3$ perovskite crystal structures [1,10-12].

As we detail below, over the decades of experimentations on the polymorphs of BaCeO$_3$, some of the differences or the disagreements between the results of experiments of different kind are reciprocally course corrected. In particular, the inconsistency between the results of Raman spectroscopy and neutron diffraction were lately reconciled and corroborated with each other [11]. However, the observations made upon the thermoanalytical experiments [12,24] differ from the number of phases and probably their sequence of structural phase transitions observed in the aforementioned neutron diffraction and Raman spectroscopy experiments and these differences remained to be uncertain till today. Here in this work, we provide plausible elucidations for the concerns raised on the crystal structures, number of phases and phase transitions by systematically studying the evolution of $BaCeO_3$ polymorphs as a function of temperature using data mining-cum-first-principles Density Functional Theoretical (DFT) calculations and phonon lattice dynamics. The resultant theoretical approach not only reproduced the crystal structure solutions and their approximate polymorphic phase transition temperatures determined based on the X-ray/neutron diffraction and polarized Raman spectroscopy experiments but also predicted a set of thermodynamically competing high temperature perovskite phases that may be attributed to the phase transitions mostly pointed in various thermoanalytical experiments. Our results provide further inferences to understand the evolution of the structures of $BaCeO_3$ at high temperature, which would also ensue in modeling appropriate structure-property correlations in building efficient high temperature $BaCeO_3$ based ion conducting solid oxide fuel cells and other electrochemical devices.

The rest of the paper is structured as follows: After the remarks on the ground state crystal structure of $BaCeO_3$, we delve into sorting experimental crystal structures of high temperature polymorphs of $BaCeO_3$, sequence of structural phase transitions and the number of phase transitions observed in various experiments in the literature. There after we present our data-mining approach coupled to first-principles theoretical calculations by which we screen the $BaCeO_3$ perovskite structural candidates. We then present the results of the predicted crystal structures and their comparative crystallographic structure systematics. Followed by this, we discuss phonon spectra and structural variations monitored with the evolution of temperature by free energy phase diagram constructed that delineate the stability of the structures and as well as the sequence of the phase transition.

**$BaCeO_3$ at STP**

The ground state crystal structure of $BaCeO_3$ has been the subject of extensive investigation ever since the compound was synthesized. In 1934, A. Hoffmann [26] initially described its crystal structure as cubic perovskite, there after time and again, a great deal of crystallographic data has been reported, but the interpretations of the structure solutions are often in disagreement as the diffraction patterns of the compound have been indexed on the basis of different crystal lattice symmetries, such as monoclinic, tetragonal, and orthorhombic unit cells. The divergence in the structure solutions is considered to be typical to perovskites in the early days of X-ray crystallography and results were only converged as the instrumentation and data-refining techniques improved with time. In his seminal account on the structural crystallography of BaCeO$_3$, Knight comprehensively reviewed the problems behind determining the correct crystal system, unit cell metric and space group of BaCeO$_3$ [10]. And, it is evident that as detailed in the review and references therein, of late several studies have confirmed that at room temperature, BaCeO$_3$ crystallizes in an orthorhombic unit cell with *Pnma* space group, which is GdFeO$_3$ structure type, whose crystal structural aspects will be discussed below along with its high temperature polymorphs. In reference to its phases at high-temperature, the orthorhombic *Pnma* is often referred as the low temperature or room temperature polymorph of BaCeO$_3$.

**High temperature BaCeO$_3$ polymorphs**

In diverse experiments, it has been shown that the BaCeO$_3$ perovskite ceramic-oxide undergoes a series of structural phase transitions between the temperature 4.2 and 1500 K. Much alike its room temperature ground state orthorhombic *Pnma* structure, the crystal structure solutions of the high temperature phases were argued upon and debated over and again. In a bipartite article in 1951, Wood has proposed a diagrammatic structure filed for polymorphism in ABO$_3$ compounds and suggested general crystallographic variations in perovskites [27]. The variations are due to unit cell elongation or compression and occupancy of anions or cations in non-special positions. At room temperature, while these variations found are predominantly related to the ones in the crystals of orthorhombic symmetry, with increasing temperature, the structures having rhombohedral, orthorhombic, tetragonal and cubic modifications (in that order) are suggested. Indeed for the case of BaCeO$_3$, except for some inconsistencies in their order of appearance over the temperature scale, the temperature induced polymorphic transformations derived from Wood's intended diagrammatic aid in has been realized in experiments.

Several research groups have studied experimentally the evolution of temperature induced polymorphic phase transitions in BaCeO$_3$. The studies have been carried out using a variety of experimental techniques. By means of X-ray diffraction as a function of temperature, Preda and Dinescu reported one of the first evidences of structural phase transitions in BaCeO$_3$ [13]. In their experiments in 1976, it was shown that BaCeO$_3$ undergoes a phase transition at 773 K from a high temperature cubic to a low temperature tetragonal structure. Such a transition was obtained by slow cooling of the sample. However, Longo et al. (1981) when subjected the specimen of BaCeO$_3$ to different cooling temperatures no evidence of the polymorphic transformation was found; the diffraction patterns of all the samples cooled with different rates were found to be identical and the near room temperature (293 K or 20 °C) data was indexed to a tetragonal unit cell [14].

Years later between 1992 and 1999, Lucazeau and co-workers performed a series of Raman spectroscopy experiments of BaCeO$_3$ and these experiments are often studied together with impedance spectroscopy and neutron diffraction experiments [11, 15-17]. In their initial set of high temperature Raman studies, Scherban et al. found that the compound BaCeO$_3$ undergoes two structural phase transitions from orthorhombic to tetragonal at 427 K and tetragonal to cubic at a much higher temperature of 1112 K [15,16]. In these experiments the symmetry of the phases identified are described based on factor group analysis and that resulted in *Pnma*, *P4/mbm*, and *Pm-3m* as the plausible crystallographic space group representations for orthorhombic, tetragonal and cubic modifications, respectively. For brevity, the proposed sequence of phase transition in BaCeO$_3$ can be written as *Pnma* → *P4/mbm* [427] → *Pm-3m* [1112], here in this present paper the space groups are denoted in Hermann-Mauguin notation and the structural phase transition temperatures in Kelvin are shown within the square brackets.

However, contrary to the results of the Raman spectroscopy experiments, Knight has reported in the year 1994 using high-resolution neutron diffraction experiments, that with increasing temperature BaCeO$_3$ undergoes three structural phase transitions from orthorhombic *Pnma* to a second orthorhombic *Imma* at 563 K that subsequently transformed to a rhombohedral *R-3c* at 673 K, and finally, the rhombohedral phase transforms into a high symmetric cubic *Pm-3m* structure at 1173 K [18]. This sequence of phase transition is succinctly written as *Pnma* → *Imma* [563] → *R-3c* [673] → *Pm-3m* [1173]. The proposed four phases of BaCeO$_3$ are crystallographically well characterized by Knight using full Rietveld analysis.

In order to understand the structural phase transitions as a function of temperature and to settle the disagreement between the results of Raman spectroscopy and neutron diffraction of BaCeO$_3$, Lucazeau and co-workers revisited the experiments. In their subsequent Raman spectroscopy experiments carried out along with impedance spectroscopy (Loridant et al.) and neutron diffraction experiments (Genet et al.) of BaCeO$_3$, complex differences are found in the temperature stability fields of the proposed perovskite structures [17, 11]. However, based on their best fits through Rietveld refinements, the phase transitions sequence and transition temperatures are shown to be reconciled with that of the results of Knight's neutron diffraction experiments [18]. The earlier proposed tetragonal *P4/mbm* phase for the *Pnma* → *P4/mbm* [427] transition and other probable tetragonal space group symmetries such as *I4/mmm* and *I4/mcm* chosen by means of Glazer classification (discussed later on), resulted in misfit in the lattice metrics of the former and large *R* factors for the later two space groups, and therefore these tetragonal phases not accepted as the best structural solutions. The appearance of tetragonal *P4/mbm* phase in the Raman studies is attributed to the pretransitional effects due to short-range local ordering of BaCeO$_3$ perovskite crystal lattice. To conclude, thus far the results of neutron diffraction and Raman spectroscopy demonstrate that the perovskite BaCeO$_3$ undergoes three structure transitions with a sequence of *Pnma* → *Imma* [563] → *R-3c* [673] → *Pm-3m* [1173].

But, of late in several thermoanalytical experiments, it has been shown that BaCeO$_3$ could exhibit a more complicated polymorphism involving more than three phase transitions [12,24]. In their dilatometry experiments, Kuzmin et al. have observed a total of nine temperatures of singularities in between 300–1300 K, which would correspond to a number of structural transitions [24]. While the more pronounced singularities noted at 520 K, 665 K, and 1140 K are in excellent agreement with the transition temperatures determined using neutron diffraction and Raman spectroscopy, there are at-least three other structure transitions rather distinctly observed at 440 K, 900 K, and 1030 K; the transition temperatures measured based on thermoanalytical experiments are to be read at best with a standard deviation of ± 10 K. This brings us to outline that there is an excellent agreement between all the experiments over the existence of a low temperature orthorhombic *Pnma* and high temperature cubic *Pm-3m* phases as the two end points in the sequence of temperature induced polymorphic phase transitions in BaCeO$_3$, however, the variability in the presence of intermediate phases (between *Pnma* and *Pm-3m*) turned-out be lies in the specificities of experiments.

Overall, a total of six structure transitions for BaCeO$_3$ are identified in various high temperature experiments, so far three (*Pnma* → *Imma* [563] → *R-3c* [673] → *Pm-3m* [1173]) transitions are well characterized with four definite crystal structures. For the remaining transitions, the structure solutions are heretofore unknown. In order, therefore to resolve the issues it is necessary excogitating the structures of BaCeO$_3$ at high temperature.

In the theory of DFT calculations and phonon lattice dynamics, one requires the knowledge of the crystal structures to investigate the evolution of BaCeO$_3$ polymorphs as a function of temperature. While crystal structure predictions may be carried out by several means, here we follow a knowledge based data mining approach, where the crystal structures mined from databases are calculated for their equilibrium geometries by minimizing the forces on the atoms in unit cell using DFT based total electronic energy crystal structure optimizations. Considering the calculated electronic energies, the crystals are ranked for plausible structural candidates and these are further investigated for their structure stability and studies of temperature induced structural phases transitions by phonon-lattice dynamics. We now present the computational methodology followed by the schemes to screen and predict the stoichiometric crystal structures to construct a free-energy phase diagram that would uncover series of phase transitions for BaCeO$_3$ as a function of temperature.

**COMPUTATIONAL METHODS**

First-principles plane wave based DFT calculations were performed to investigate the electronic structure and thermodynamic properties of BaCeO$_3$ polymorphs. For all the calculations, the description of exchange-correlation energy functional of Perdew, Burke, and Ernzerhof (PBE) and the projected augmented wave (PAW) pseudopotentials were used, as implemented in the Vienna Ab initio Simulation Package (VASP) program [28-31]. For the elements in the subject compound, the pseudopotentials consisting the valence electron configurations were chosen as Ba:($5s^25p^66s^2$), Ce:($5s^25p^64f^15d^16s^2$), and O:($2s^22p^4$). For initial screening of all the structures by total electronic energies, a plane-wave basis with kinetic energy cut-off of 520 eV and for the most relevant stable structures identified a more accurate cut-off of 800 eV was applied. All the results presented here in the subsequent sections are from the calculations performed with the kinetic energy cut-off of 800 eV. The Brillouin zone (BZ) sampling was done using the automatic k-mesh generation scheme with k-point grids of 2π × 0.0125 Å$^{-1}$ and 2π × 0.025 Å$^{-1}$ for primitive and supercell calculations, respectively. A self-consistence convergence of $10^{-7}$ eV was set for the electronic energy minimizations. During the crystal structure geometry optimizations of BaCeO$_3$ polymorphs, the forces acting on the ions were minimized up to the $10^{-3}$ eV/Å, where the lattice parameters, cell volume, and ionic positions were fully relaxed without applying any geometric constraints.

To calculate the structure stabilities and free energies of the proposed BaCeO$_3$ polymorphs, phonon lattice dynamics were carried out within the harmonic approximation, applying density functional perturbation theory (DFPT) and as well as finite displacement methods [32, 33]. PHONOPY interfaced with VASP was used to calculate the necessary forces and the force constants required constructing the phonon spectra through out the zone and to estimate the thermodynamic properties [34]. For all the BaCeO$_3$ polymorphs in their ground states, the free energies were estimated by

$$F(V_0, T) = E_{el}(V_0) + F_{ph}(V_0, T) \text{ -------- (1)}$$

where $E_{el}$ is the total electronic energy calculated per unit cell at T= 0 K, and $F_{ph}(V_0, T)$ is the vibrational contribution to the free energy (Helmholtz) and it can be obtain from

$$F_{ph}(V_0, T) = \frac{1}{2}\sum_{q,v} \hbar\omega_{q,v} + k_B T \sum_{q,v} \ln\left[1 - \exp\left(-\frac{\hbar\omega_{q,v}}{k_B T}\right)\right] \text{ -------- (2)}$$

where $\omega_{q,v}$ is the phonon frequency at wave vector $q$ and $v$ is band index. T, $k_B$, and $\hbar$ are temperature, Boltzmann constant, and Planck constant, respectively. The first term in equation (2) is the zero-point vibrational energy and the second term consists of each vibrational mode's contribution due to the thermal occupation of phonons modes. The sums are taken over all the phonon branches at each $q$ of the Brillouin zone of the crystal lattice.

## RESULTS AND DISCUSSIONS
### Data mining-cum-DFT

The crystal structures are data mined by making use of the ICSD and Pauling Files inorganic crystal structure databases [35-38]. Several polytypes are systematically mined by screening for ABX$_3$ perovskite composition, where A and B represent the cations with 2+ and 4+, whereas the element X is for anion with 2- formal oxidation states, respectively. Considering these elemental formal oxidation states, we chose the following filters to screen the structures for ABX$_3$.
a) For A-site elements: Alkali-earth metals, IV-group, transition metal series, lanthanide and actinides are considered.
b) For B-site elements: IV-group, transition series, lanthanide and actinides are considered.
c) For the anionic X-site: Elements from VI-group are selected.

The transition metals, lanthanide and actinides are well known for their multiple valencies, and therefore in our search criteria, we consider these elements for A and B cations. As a result of imposing these rules in mining the crystal structures for BaCeO$_3$, we have found a total of 61 crystal structures that are distinctly different from each other in terms of their crystallographic space groups. The structures in this collection are then colored with Ba, Ce and O for the A, B and X-type elements, respectively. Upon crystal structure optimizations as described in the computational methods, the 61 structures are further reduced to 34 as some of the structures are converged to the same symmetry that of the others in the initial collection. In our workflows, we have adopted findsym routines to carefully identify the crystallographic space groups of the ground state crystal structures obtained through DFT calculations [39,40]. The 34 distinctly different crystal structures of BaCeO$_3$ are further evaluated for ranking the structures based on the total electronic energy calculated at T= 0 K and about half of these sorted with respect to the lowest energy structure with increasing energy order are shown in Fig. 1.

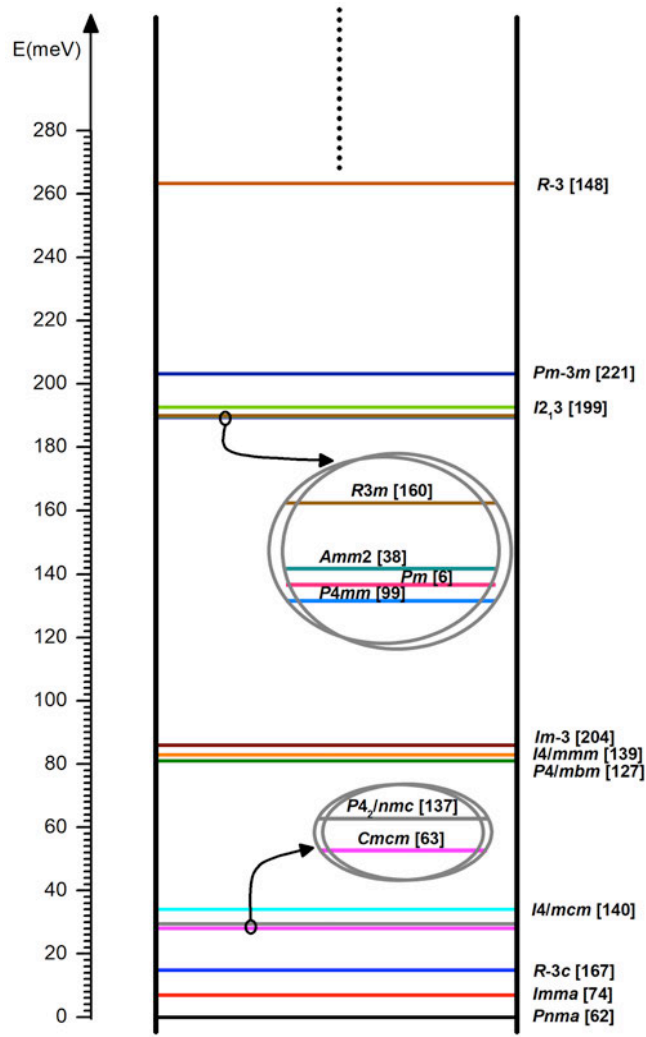

FIG. 1. Energy ladder diagram that illustrates the ordering of calculated ground state electronic energies of the 16 lowest energy crystal structures of BaCeO$_3$ predicted by means of data-mining-cum-DFT calculations. The electronic energy of each structure is designated by a horizontal rung between the vertical rails of the ladder. Each phase is labeled by its space group symbol and the corresponding space group number is listed in the bracket. The energies are shown with reference to the most stable polymorphs of BaCeO$_3$ (*Pnma*). Structures that are energetically close to degenerate are shown in zoomed-in insets.

From Fig. 1, it is remarkable that the best 15 in 34 contains all the four experimentally well-characterized $BaCeO_3$ polymorphs and 11 new phases. On increasing energy scale, the first three lowest energy structures: orthorhombic *Pnma*, orthorhombic *Imma*, and rhombohedral *R-3c* (in that order) are turned-out to be the well known phases that are experimentally identified by Knight in probing the sequence of *Pnma* → *Imma* [563] → *R-3c* [673] → *Pm-3m* [1173] temperature induced phase transitions in $BaCeO_3$ [18]. While the orthorhombic *Pnma* is the most stable polymorphs of $BaCeO_3$, the *Imma* and *R-3c* phases are calculated to be less stable by 7 meV and 15 meV, respectively, all with reference to the *Pnma*, per formula unit. The other experimentally known cubic *Pm-3m* phase is placed at its $15^{th}$ position with relative energy of 203 meV. In other words, it is to say that in our calculations at T= 0 K, there exist at-least 11 polytypes as stationary points between the experimentally known rhombohedral *R-3c* and cubic *Pm-3m* phases.

Interestingly, in 11, we find a tetragonal phase *P4/mbm* at $7^{th}$ rank (122 meV lower to the cubic *Pm-3m*) that was initially deduced by Lucazeau and co-workers by means of Raman spectroscopic measurements [15-17]. In addition to this, one of the novel phases that we have identified is an orthorhombic *Cmcm*, placed at $4^{th}$ rank, which is just about 13 meV less preferred to *R-3c* phase or 175 meV (or about 4 kcal/mol) more stable to the cubic *Pm-3m* phase. We note that, Knight and Bonanos considered the *Cmcm* phase as one of the possible space groups for $BaCe_{1-x}Nd_xO_{3-\delta}$ (x = 0.2), but the Rietveld refinements were proven to be unsuccessful [41]. The *Cmcm* is one of the space groups listed in the Glazer classification of perovskite tilt system [42,43]. The other probable space groups, *I4/mmm* and *I4/mcm* chosen by Genet et al. [11], (again under Glazer's tilt system) for the intermediate temperature $BaCeO_3$, but are then rejected due to large *R* factors relative to *R-3c* also resulted in our calculations as the thermodynamically alternative competing phases (see Fig. 1). A tetragonal phase with space group $P4_2/nmc$ that hasn't been experimentally assigned so far to $BaCeO_3$ is turned out be almost thermodynamically degenerate to the *I4/mcm* phase. Nonetheless, we find that, in between *Cmcm*, $P4_2/nmc$, *I4/mcm*, *P4/mbm*, *I4/mmm* and *R-3c*, the later is found be thermodynamically the most competing polymorphs of $BaCeO_3$ at T= 0 K, which correlate well to the fact that the calculated thermodynamic stability order *Pnma* > *Imma* > *R-3c* at T= 0 K comparable to the known experimental sequence of temperature induced phase transitions determined by neutron diffraction and Raman spectroscopic investigations [10,11]. But, as noted, there are 11 calculated phases between rhombohedral *R-3c* and cubic *Pm-3m*; the later is known to be the highest temperature polymorph of $BaCeO_3$. It is also interesting to mention that the calculations predict a different cubic phase (*Im-3*) that is 117 meV more stable ($9^{th}$ rank) than the known cubic *Pm-3m* structure ($15^{th}$ rank). As we shall show that the theoretically predicted phases at T= 0 K are found to be important in understanding the prediction of the phase transitions in $BaCeO_3$ as a function of temperature. We now shall discuss the crystal structures of thermodynamically the most stable phases predicted (Fig. 1) and as well as some of the important structures found relevant to understand the evolution of temperature driven phase transitions in $BaCeO_3$.

**Crystal structural aspects**

For the crystal structure analysis and construction of the phase diagram, we have selected a total of ten lowest energy structures. But, as the cubic *Pm-3m* phase is experimentally known, it has also been included in the selected ten for comparison, although it stands at 15$^{th}$ rank. The other important structures that are energetically more preferred than the cubic *Pm-3m* phase are given in the Supplemental Material of this paper [44], Fig. S1. Based on the crystal system, the ten structures are classified into four types: (1) Orthorhombic: *Pnma* (No. 62), *Cmcm* (No. 63) and *Imma* (No. 74). (2) Tetragonal: *P4/mbm* (No. 127), *P4$_2$/nmc* (No. 137), *I4/mmm* (No. 139) and *I4/mcm* (No. 140). (3) Trigonal: *R-3c* (No. 167). (4) Cubic: *Im-3* (No. 204) and *Pm-3m* (No. 221). The ground state crystal structures of these ten best structural candidates are shown in Fig. 2, in the order of their structure stabilities at T= 0 K.

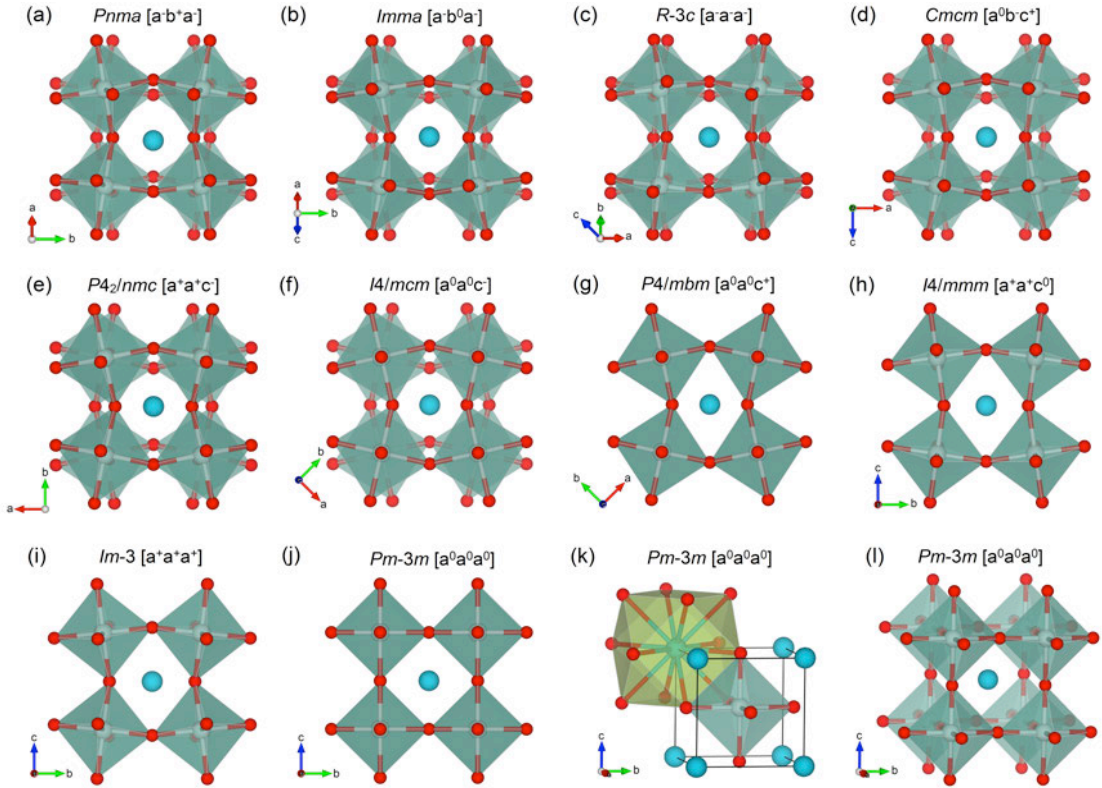

FIG. 2. Calculated ground state crystal structures of the ten lowest energy polymorphs of BaCeO$_3$: (a) *Pnma* (62), (b) *Imma* (74), (c) *R-3c* (167), (d) *Cmcm* (63), (e) *P4$_2$/nmc* (137), (f) *I4/mcm* (140), (g) *P4/mbm* (127), (h) *I4/mmm* (139), (i) *Im-3* (204) and (j) *Pm-3m* (221). The Glazer octahedral tilt system and the Hermann-Mauguin (H-M) space group notation are used in parentheses to identify each polymorph. (k) Polyhedral representations of [CeO$_6$] octahedral and [BaO$_{12}$] cuboctahedron network formation in the cubic perovskite. (l) Cubic perovskite structure same as (j) from a different view. Ba, Ce, and O atoms are depicted in cyan, grey, and red color, respectively. The orientation of the crystal axes (*a*, *b*, and *c*) is displayed using a compass. The crystal structures are visualized using VESTA [57].

The structure of BaCeO$_3$, in its ground state, adopts an orthorhombic *Pnma* hettotype crystal structure (Fig. 2a), which is a distorted variant of an aristotype cubic (*Pm-3m*) perovskite (Fig. 2j). In a family of crystal structures, the word aristotype refers to the highest symmetric crystal structure, and the hettotype is for lower symmetric analogues [45]. In fact, all the structures predicted here are of hettotype and these can be considered as the derivatives of the aristotype cubic perovskite. The distortions are of the nature of tilting and rotations of $CeO_6$ octahedral units (see the different tilts with respect to the crystallographic compass shown in Fig. 2 and the discussion below), which leads to the reduction in the symmetry of the resultant crystal lattices. These distortions and in general the symmetry relations between the crystal lattices of the perovskites are described by an octahedral tilt system. While there could be innumerable distortions, the crystallographic principles-cum-group theoretical relations constraints the octahedral tilts to a certain number of space groups. For $ABX_3$ perovskites, considering the cell doubling, the tilt system of Glazer and its extended versions based on crystallographic aspects (the tilt systems of Aleksandrov, Howard-Stokes, and Gopalan-Litvin) have shown that the number of tilt systems should be limited to 15 distinctly different space groups (excluding the combinations, otherwise a maximum of 23 tilt systems) [46-48]. Interestingly in our data mining approach of predicting plausible polymorphs for $BaCeO_3$, all the first ten lowest energy structural candidates (Fig. 1 and 2, including the cubic *Pm*-3*m*) are found to have only the space groups evaluated by the tilt systems. The other high-energy phases are predominantly found to be in non-perovskite like structure systems. All the predicted crystal structure coordinates of $BaCeO_3$ polymorphs are listed in the Supplemental Material [44], Appendix SA1.

Using the octahedral tilt system, for structural comparisons to be made across and to bring in the crystallographic aspects of all the perovskites studied here, we begin to consider the structure of $BaCeO_3$ *Pm*-3*m* cubic perovskite. In this structure, one may view from Fig. (2k) that the $Ce^{4+}$ cation occupies the center of the unit-cell in the cubic system, which is octahedrally (6-fold) coordinated to the $O^{2-}$ anions at the face centers, forming $CeO_6$ octahedron that in turn forms a vertex sharing three-dimensional extended octahedral network. Such $CeO_6$ octahedral network generates voids at the cube vertices. The voids are occupied by the $Ba^{2+}$ cation that can have 12-fold coordination with the $O^{2-}$ anions, forming $BaO_{12}$ cuboctahedron. The perovskite thus forming also can be visualized by shifting its center to the origin of the unit cell, where the $CeO_6$ octahedral units run along the orthogonal axis of the cubic lattice vectors by corner sharing [Fig. (2l)] and the $Ba^{2+}$ cation occupies the body-centered void of the cube generated by the $CeO_6$ octahedral network.

As one would see from Fig. 2a-2i, indeed, the structures are distorted variants of the cubic perovskite (Fig. 2j) and these distortions are due to tilting of $CeO_6$ octahedral units. In the Glazer's classification of the perovskites [42,43], the tilting of an octahedron in terms of component tilts is considered along the three-tetrad axes (face normals) of the cube. In successive layers of octahedra, the magnitudes and directions of the tilts about the triad [100], [010] and [001] are denoted by $a^l$, $b^m$, and $c^n$ respectively. The letters (*a*, *b*, and *c*) refer to the magnitude of the tilt in degrees and the superscripts are for the direction of the tilt, that is along certain axis, the tilts could be either in-phase (+) or anti-phase (−) or can even have no tilt (0). Note that the letters **a**, **b**, and **c** are repeated the same for equal angle of tilts, for example, $a^0a^0a^0$ represents the aristotype *Pm-3m* cubic perovskite with no tilts along [100], [010], and [001], respectively [Fig. (2j)]. Similarly, $a^+a^+a^+$ represents equal tilts but all with in-phase tilts, which corresponds to the 9th rank hettotype cubic structure (*Im-3*) shown in Fig. (2i). One more, for instance, in the context of the structures that we predicted here, $a^0b^-c^+$ is a hettotype orthorhombic (*Cmcm*) perovskite that indicates all the tilts are unequal in magnitudes but with no tilt along [100], and the tilts along [010] and [001] are anti-phase and in-phase, respectively [Fig. (2d)]. The most stable BaCeO$_3$ perovskite goes by one of the most commonly occurred tilt systems of perovskite, orthorhombic *Pnma* with $a^-b^+a^-$ tilt system [Fig. (2a)], where the CeO$_6$ octahedral units are stabilized with anti-phase tilts of equal magnitude about the *a* and *c*-axes that is along [100] and [001], respectively. However, viewing down the *b*-axis the successive CeO$_6$ octahedra are tilted in-phase with a different tilt angle than that of the tilts along the *a* and *c*-axes.

The tilt systems calculated and the computed ground state crystal structure data — lattice parameters, unit-cell volumes, (Ce/Ba)-O distances, and Ce-O-Ce tilt angles of all the phases of BaCeO$_3$ shown in Fig. 2 are reported in Table 1. For comparison, the experimental data is also provided. The calculated crystal structure geometric parameters corroborate well with the experimental values [18,49]. For example, compared to the experiment, in particular for the low temperature phase (*Pnma*), the calculated lattice parameters and unit cell volumes are overestimated by about 1% and 3.6% respectively. And, these deviations are well within the mean relative errors (about 5% in volumes) estimated for various solid-state materials by DFT calculations [50-52]. However, considering the high temperature orthorhombic *Imma*, rhombohedral *R-3c* and cubic *Pm-3m* phases, the calculated volumes are only varied by about 2% to the experimental values. That somewhat better agreement is understandable, because the experimental volumes that are increased by about 4% due to thermal expansion are compared to the volumes calculated at T= 0 K. Similar variations are noted for Ba-O or Ce-O distances and Ce-O-Ce angles as tabulated in Table 1. This gives us confidence that when the structures predicted here are observed, similar comparisons should hold true.

**Phonon lattice dynamics**

Phonon calculations are performed for all the ten structures shown in Fig. 2. As such, since the phonon calculations demand huge computational resources, in particular several polymorphs that we have with varied cell dimensions, extensive initial screening of the dynamic structure stabilities and temperature induced phase transitions are carried-out using various supercells and phonon modulations (see Figs S2 and Table S1 in the Supplemental Material [44]). On the basis of the number of novel structure transitions yielded and experimental structures in total, we selected six of the ten structures for a more detailed study of phonons, the results of which are presented here. The phonons for the remaining phases are given in the Supplemental Material [44], Figs S3. The results are striking, the selected six phases of BaCeO$_3$ consists of the four well-known experimental crystal structures (orthorhombic *Pnma* and *Imma*, rhombohedral *R-3c*, and cubic *Pm-3m*) and two novel phases (orthorhombic *Cmcm* and tetragonal *P4/mbm*) predicted by us through data driven approaches found to be competently stabilized with increasing temperature, marking two well defined critical temperatures. We now first present the analysis of the phonons for the six phases and proceeded by the construction of free energy phase diagram to study the temperature induced phase transitions in BaCeO$_3$ polymorphs.

TABLE I. The calculated crystallographic information of BaCeO$_3$ polymorphs: The results of the calculations are compared with the available experimental data. Crystallographic space group symbols, structure types, and Glazer tilt system are used to denote the polymorphs. Z-value refers to the number of formula units present in a unit cell.

| Structural parameters | *Pnma* (62) | | *Imma* (74) | | *R-3c* (167) | | *Cmcm* (63) |
|---|---|---|---|---|---|---|---|
| | Theory | Experiment | Theory | Experiment | Theory | Experiment | Theory |
| Structure-type | GdFeO$_3$ | GdFeO$_3$ | BaPbO$_3$ | BaPbO$_3$ | LaAlO$_3$ | LaAlO$_3$ | NaNbO$_3$ |
| Lattice parameter | | | | | | | |
| a (Å) | 6.292 | 6.209 | 6.273 | 6.233 | 6.259 | 6.243 | 8.845 |
| b (Å) | 8.869 | 8.760 | 8.851 | 8.795 | 6.259 | 6.243 | 8.941 |
| c (Å) | 6.288 | 6.221 | 6.323 | 6.262 | 6.259 | 6.243 | 8.884 |
| α (°) | 90.0 | 90.0 | 90.0 | 90.0 | 60.6 | 60.2 | 90.0 |
| β (°) | 90.0 | 90.0 | 90.0 | 90.0 | 60.6 | 60.2 | 90.0 |
| γ (°) | 90.0 | 90.0 | 90.0 | 90.0 | 60.6 | 60.2 | 90.0 |
| Volume (Å$^3$/f.u.) | 87.73 | 84.61 | 87.77 | 85.83 | 87.84 | 86.39 | 87.83 |
| Z-value | 4 | 4 | 4 | 4 | 2 | 2 | 8 |
| Avg. bond length | | | | | | | |
| Ba-O (Å) | 3.162 | 3.124 | 3.117 | 3.130 | 3.158 | 3.135 | 3.158 |
| Ce-O (Å) | 2.269 | 2.242 | 2.266 | 2.235 | 2.264 | 2.235 | 2.263 |
| Bond angle (°) | | | | | | | |
| Ce-O-Ce | 157.5 | 157.9 | 159.4 | 163.5 | 158.0 | 162.9 | 154.1 |
| | 154.9 | 154.7 | 153.9 | 157.9 | 158.0 | 162.9 | 164.6 |
| | 157.5 | 157.2 | 159.4 | 163.5 | 158.0 | 162.9 | 157.2 |
| Tilt system | a$^-$b$^+$a$^-$ | a$^-$b$^+$a$^-$ | a$^-$b$^0$a$^-$ | a$^-$b$^0$a$^-$ | a$^-$a$^-$a$^-$ | a$^-$a$^-$a$^-$ | a$^0$b$^-$c$^+$ |

Table 1 continued …..

| Structural parameters | *P4$_2$/nmc* (137) | *I4/mcm* (140) | *P4/mbm* (127) | *I4/mmm* (139) | *Im-3* (204) | *Pm-3m* (221) | |
|---|---|---|---|---|---|---|---|
| | Theory | Theory | Theory | Theory | Theory | Theory | Experiment |
| Structure-type | CaSiO$_3$ | SrZrO$_3$ | NaNbO$_3$ | CaSiO$_3$ | CaSiO$_3$ | CaTiO$_3$ | CaTiO$_3$ |
| Lattice parameter | | | | | | | |
| a (Å) | 8.858 | 6.244 | 6.246 | 8.931 | 8.911 | 4.474 | 4.445 |
| b (Å) | 8.858 | 6.244 | 6.246 | 8.931 | 8.911 | 4.474 | 4.445 |
| c (Å) | 8.955 | 9.003 | 4.515 | 8.858 | 8.911 | 4.474 | 4.445 |
| α (°) | 90.0 | 90.0 | 90.0 | 90.0 | 90.0 | 90.0 | 90.0 |
| β (°) | 90.0 | 90.0 | 90.0 | 90.0 | 90.0 | 90.0 | 90.0 |
| γ (°) | 90.0 | 90.0 | 90.0 | 90.0 | 90.0 | 90.0 | 90.0 |
| Volume (Å$^3$/f.u.) | 87.84 | 87.76 | 88.08 | 88.33 | 88.44 | 89.57 | 87.81 |
| Z-value | 8 | 4 | 2 | 8 | 8 | 1 | 1 |
| Avg. bond length | | | | | | | |
| Ba-O (Å) | 3.155 | 3.156 | 3.159 | 3.163 | 3.156 | 3.164 | 3.143 |
| Ce-O (Å) | 2.263 | 2.260 | 2.259 | 2.259 | 2.259 | 2.237 | 2.222 |
| Bond angle (°) | | | | | | | |
| Ce-O-Ce | 155.0 (155.6) | 154.2 | 155.7 | 162.4 | 161.0 | 180.0 | 180.0 |
| | 155.0 (155.6) | 154.2 | 155.7 | 162.4 | 161.0 | 180.0 | 180.0 |
| | 166.7 | 180.0 | 180.0 | 157.2 | 161.0 | 180.0 | 180.0 |
| Tilt system | a$^+$a$^+$c$^-$ | a$^0$a$^0$c$^-$ | a$^0$a$^0$c$^+$ | a$^+$a$^+$c$^0$ | a$^+$a$^+$a$^+$ | a$^0$a$^0$a$^0$ | a$^0$a$^0$a$^0$ |

The calculated phonon dispersion spectra and phonon densities of states (DOS) of the selected six phases of $BaCeO_3$ are presented in Fig. 3(a-f) and 3(g-l), respectively; phonon dispersion plots with extended $q$-path are shown in in the Supplemental Material [44], Fig. S3. From the presented phonon spectra, it is evident that the low temperature orthorhombic *Pnma* comprises all the phonon branches with real frequencies, that is to say that the structure has no imaginary frequencies throughout the Brillouin zone, which indicate the dynamical stability of the structure. The phonons calculated for the remaining structures [Fig 3(b–f)] are found to have imaginary frequencies and therefore, these phases should be regarded as dynamically unstable at low temperatures (T → 0 K) and these may be considered as the high-temperature structural candidates that are not quenchable down to low temperatures. Indeed, as per as several experiments on structural phase transitions in $BaCeO_3$ are concern and as we shall also show in the section below that the free-energy phase diagram calculated together reveals that, on cooling the supposed high temperature phases spontaneously transformed to low temperature orthorhombic *Pnma* structure.

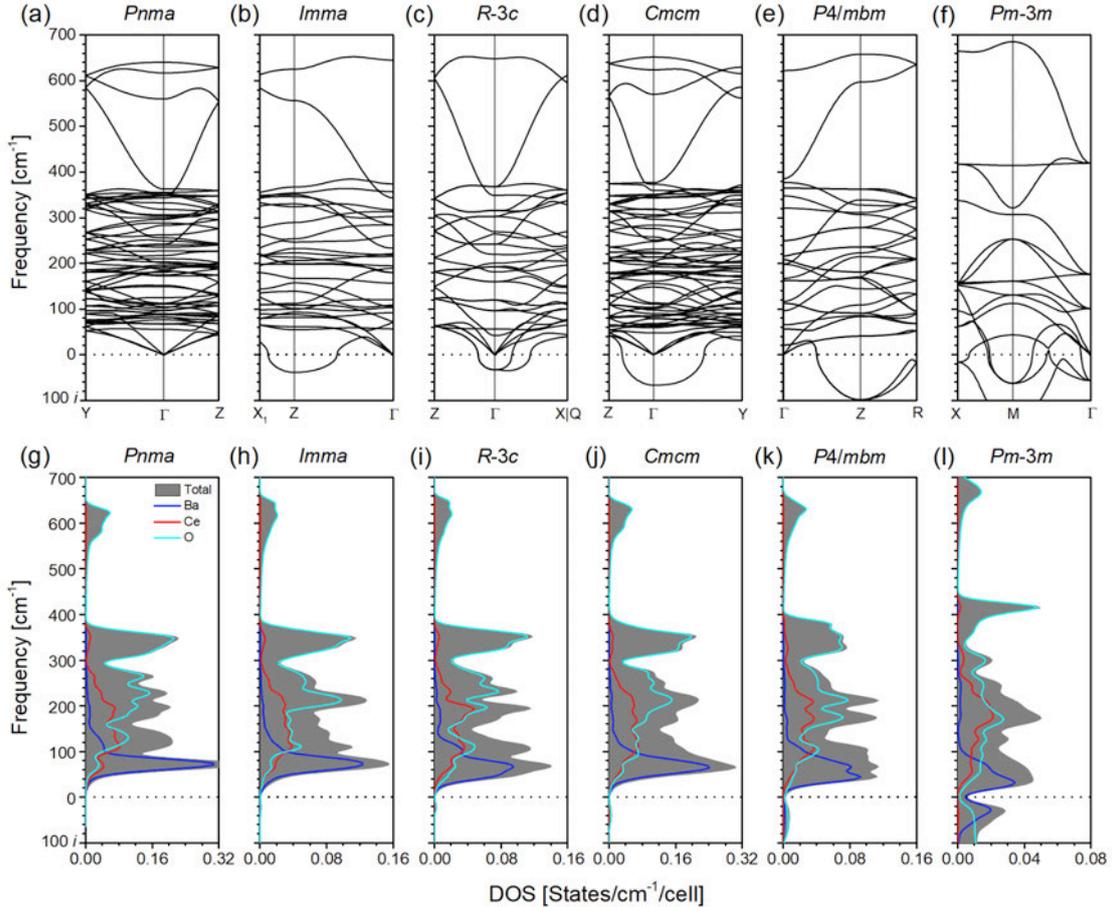

FIG. 3. (a)−(f) Calculated phonon band dispersion spectra at certain high symmetry points in the first Brillouin zone and (g)−(l) calculated phonon density of states of the elected six $BaCeO_3$ polymorphs. The total phonon density of states are shown with the shaded region, whereas the atom projected phonon density of states for Ba, Ce, and O atoms are shown in blue, red, and cyan colors, respectively.

As can be seen from the DOS in Fig. 3(g-l), the lightest element in $BaCeO_3$, the oxygen atom, contributes to the highest phonon frequency region, although predominantly between 500 $cm^{-1}$ and 660 $cm^{-1}$; its contribution spans across the entire phonon spectra coupling to the phonon segments of the cations presented. The oxygen vibrational coupling with cerium is substantial than to Barium, as besides the differences in their atomic mass, the internuclear separation of Ce-O is about 30% shorter than Ba-O (see Table 1). Barium being the heaviest element here, it contributes to the lowest frequency region, from ~0 $cm^{-1}$ to 150 $cm^{-1}$. And, above which cerium take part until reaching an average maximum value of 400 $cm^{-1}$. As discussed earlier, barring the structure of *Pnma*, the other phases feature imaginary frequencies. The magnitudes of the largest imaginary phonon modes in *Imma* (39$i$ $cm^{-1}$), *R-3c* (33$i$ $cm^{-1}$), and *Cmcm* (66$i$ $cm^{-1}$) phases of $BaCeO_3$ are relatively small compared to the *P4/mbm* (98$i$ $cm^{-1}$) and *Pm-3m* (150$i$ $cm^{-1}$) phases. These modes are located at the high symmetry Z (½, ½, −½), Γ (0, 0, 0), Γ (0, 0, 0), Z (0, 0, ½), and M/R (½, ½, 0)/((½, ½, ½) points of the Brillouin zone, respectively, as shown in Fig 3(b-f) and as well as in Fig. S3. The corresponding eigenvectors of the aforementioned imaginary modes are shown in Fig. 4.

It is interesting to note that, the eigenvectors of all the imaginary modes adduced chiefly feature characteristic oxygen displacements that are responsible for various $CeO_6$ octahedral tilts discussed in the previous sections (also see Fig. 2). Nonetheless, the imaginary modes propagating through zone points are rather shallow and soft, translating almost negligible densities of states for the phases of *Imma*, *R-3c*, and *Cmcm*. But for *P4/mbm* and *Pm-3m* phases, the imaginary modes are degenerate and that reflect a significant structure in the DOS as shown in Fig. 3(k and l). In order to verify whether, when the eigenvectors of the imaginary modes are applied to the crystal structures produce any new phase that would be dynamically stable at T=0 K, we have systematically followed all the eigenvectors of the imaginary modes shown in Fig. 4.

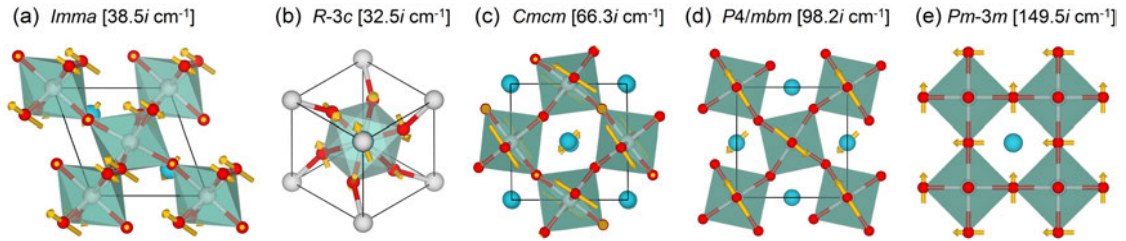

FIG. 4. Eigenvectors of the vibrational modes corresponding to the highest imaginary frequency located at specific high symmetry points of the first Brillouin zone (see text and Fig. 3 for more description) of $BaCeO_3$ polymorphs. The oxygen displacements feature librational motions. The eigenvectors and the Ba, Ce, and O atoms are shown in yellow, blue, red, and cyan colors, respectively. The magnitude of the imaginary frequency (eigenvalue) for each structure shown is given in the parenthesis.

Appropriate unit cells are constructed by following the eigenvectors of the imaginary modes and these in turn are fully optimized – all such attempts of all the modulated structures finally converged to the low temperature orthorhombic *Pnma*, which manifest the fact that structures found with imaginary modes are not recoverable as stable structures at low temperatures (T → 0 K). A detailed account on the phonon modulated structures obtained by following the imaginary modes of the eigenvector are given in the Table S1 presented in Supplemental Material. For example, the high temperature rhombohedral *R-3c* structure transformed to low temperature orthorhombic *Pnma* via the structure transformations of *R-3c* → $P\bar{1}$ → *Imma* → *Pnma,* a result that is in agreement with the recent calculations of Stoffel and Dronskowski [53], indicated that the experimentally known *Imma* is the intermediate between the phases of *Pnma* and *R-3c*. The other structures that we have predicted here through data mining, the tetragonal *P4/mbm* and the orthorhombic *Cmcm* transformed as *P4/mbm* or *Cmcm* → *P2$_1$/m* → *Pnma*. That way we predict here that on cooling, the high temperature BaCeO$_3$ phases would spontaneously transform to the low temperature orthorhombic *Pnma* structure via multiple intermediate phases. This our theoretical observation already vindicate that there could be many transient intermediate phases than the three high temperature phases observed in neutron diffraction studies. The calculated free energies of the phases that we discuss below support this proposition.

**Structural phase transitions**

To investigate the occurrence of phase transitions in BaCeO$_3$ as a function of temperature, we have constructed the free-energy phase diagram. As we shall reckon through inferred estimates (see Fig S2 in the Supplemental Material) [44], we take into account the effect of thermal expansion on change in the free energy to be compared proportionately with the calculated sequence of phase transition to the one observed in experiments [10-12]. It has been estimated that just the volume of the *R-3c* phase has to be improved by at-least 6% relative to that calculated at T= 0 K (see Table 1). We call this volume expanded variant of *R-3c* phase as *R-3c'*. Fig. 5 shows the calculated free energy vs. temperature phase diagram of BaCeO$_3$ polymorphs that include the ten best structures shown in Fig. 2 and as well as for the analysis and interpretation of the calculated free energy phase diagram to be in inconformity with the experimental observations, we consider the *R-3c'* phase.

The calculated free energies depicted in Fig. 5 are with reference to the *Pnma*, where the free energies it-self are calculated according to the equation (1) and (2). At T= 0 K, the free energy includes the vibrational zero point energy (ZPE); at this temperature the stability order of the polymorphs is found to be the same as per as the phases presented in the electronic energy ladder diagram (Fig. 1). That is, to imply that, the inclusion of the ZPE dynamical correction did not alter the stability order of the BaCeO$_3$ polymorphs.

As evident from Fig. 5 that, with respect to *Pnma*, all the phases, except more so the cubic *Pm-3m*, are increasingly stabilized under temperature, featuring four critical points at 556 K, 558 K, 666 K, and 1210 K for the transitions *Pnma* → *Imma*, *Imma* →*R-3c'*, *R-3c'* → *Cmcm,* and *Cmcm* → *P4/mbm*, respectively. The first transition obtained in our calculation at 556 K for the *Pnma* → *Imma* is in good agreement with experimental observed value of 563 K, whereas the second transition temperature, at 558 K for the *Imma* → *R-3c'* is quite under estimated to the experimental value of 673 K; it is 216 K for *Pnma* → *R-3c*. The new phases predicted here *Cmcm* and *P4/mbm* supersede *Pnma* at 666 K and 1210 K, respectively. The aforementioned transition temperatures may be attributed to the spikes observed in dilatometry at 900 K and 1030 K, respectively [12,24]. The relative trends in the phase transitions obtained in our harmonic phonon calculations although agree well with experiments, the magnitudes of the calculated transition temperatures differ, indicating the importance of anharmonic corrections, may be akin to the case of strongly anharmonic halide perovskite, where beyond first order self-consistent phonon theory (SC1) such as the accurate quasi particle (QP) based phonon calculations only recently could be able to predict the correct sequence and phase transition temperature (α-$CsPbBr_3$: cubic → tetragonal) [54]. $BaCeO_3$ certainly would be another good test case for accurate QP models. However, inasmuch as these harmonic modes affect thermodynamic stability fields, the free energy phase diagram constructed is already remarkable, featuring several critical points.

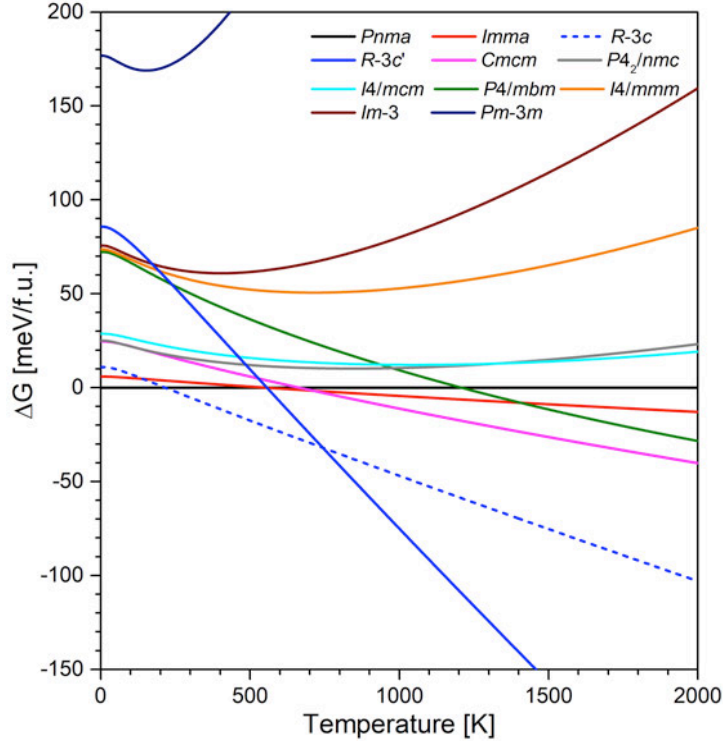

FIG. 5. Calculated free energy (ΔG)–temperature plot for the ten lowest energy polymorphs of $BaCeO_3$ and the 6% volume expanded R-3c′ phase relative to the ground state *Pnma* phase.

The other probable tetragonal phases (*P4$_2$/nmc* and *I4/mcm*) assumed for the intermediate temperature stability field appeared to be quite competing to *Pnma*, but no structure transitions have been found until reaching 2000 K; the variations in their free energies are rather flat and about a maximum of 30 meV less preferred to the *Pnma*. Interestingly, based on the gradual tilting rule [55] and anelastic mechanical behavior [56], it has been reported that the *I4/mcm* (monodomain) may exist as an intermediate phase between *Pnma* and *Pm-3m* phases. The $BaCeO_3$ structures with space groups *I4/mmm* and *Im*-3 destabilize with increasing temperature. While none of these new phases implied in either X-ray/neutron diffraction or the experiments based on polarized Raman spectroscopy, the thermoanalytical experiments indicated appearance of several phases to which these may be attributed. However, the novel high temperature structures predicted here would be fleetingly metastable, as the rhombohedral *R*-3*c* phase is found to be the most stable polymorph for a large temperature stability field; otherwise, they should proceed as stable points in the sequence of temperature induced structural phase transitions in $BaCeO_3$ perovskite.

**CONCLUSIONS**

In summary, temperature induced polymorphism in $BaCeO_3$ perovskite has been investigated by means of crystal structure data-mining-cum-first principles electronic structure and phonon lattice dynamics. Out of several stationary points predicted on the potential energy landscape of $BaCeO_3$, a total of ten best structures have been thoroughly investigated to construct the free-energy phase diagram. It is interesting that the most stable structures predicted for high-temperature $BaCeO_3$ follow the Glazer classification of perovskite tilt system. In that pool, while the calculated order of polymorphic structure stability is found to be in reasonable agreement with the neutron diffraction or polarized Raman spectroscopy based experimental sequence (*Pnma* → *Imma* → *R*-3*c*' → *Pm*-3*m*) of phase transitions, we have predicted a number of high temperature novel polymorphs that are thermodynamically competing to *Pnma*. The phases *Cmcm* and *P4/mbm* supersede *Pnma* at 666 K and 1210 K, respectively, and *P4₂/nmc* and *I4/mcm* are only within 20 to 30 meV (equivalent to 230 to 350 K) reach throughout the temperature (0 to 2000 K) studied, indicating at-least these would be fleetingly metastable at high-temperature. Such phases may have been causative for the appearance of multiple temperature singularities observed in the thermoanalytical experiments.

For the intermediate temperature (between *Pnma* and *Pm*-3*m*), among all the phases predicted in our calculations, the rhombohedral *R*-3*c* $BaCeO_3$ is found to be the most stable polymorph with a broad temperature stability field, therefore masking the existence of every other polymorph until may be reaching the cubic phase. This emphasizes the sharp appearance of *R*-3*c* phase with only a first order transition in neutron diffraction experiments and every other transition are of second order or displacive in nature. The phonon calculations carried-out for the ground state equilibrium structures underscores this observation that the high-temperature phases are at best stable only in their respective observed stability fields, but are non quenchable, on cooling down to room temperature, where upon following the eigenvectors of the soft-phonon imaginary modes that are of typical to oxygen displacements, reduced to the ground state low temperature orthorhombic *Pnma*. We propose that doping $BaCeO_3$ would be one of the routes for quenching its high-temperature phases or even to more stabilize $BaCeO_3$ at high temperature that would avoid incidents of the most fleetingly metastable phases. Such structure stability correlations would offer engineering better materials for fuel cell electrochemical devices.

ACKNOWLEDGMENTS

We acknowledge the support and the resources provided by IIT Kanpur Computer Centre High Performance Computing facility (HPC2010 and HPC2013), PARAM Sanganak under the National Supercomputing Mission (NSM) Government of India, CHMHPC at the Department of Chemistry and HPC/RNJJ through Initiation Grant No. IITK/CHM/20130116. F.N., M.S.C., K.Y., and S.S. thank the Council of Scientific and Industrial research (CSIR), Government of India, for the SRF research fellowship. A.K. thanks IIT Kanpur for his doctoral fellowship. D.L.V.K.P. acknowledges IIT Kanpur for financial support.